# Defect entropies and enthalpies in Barium Fluoride

**Vassiliki Katsika-Tsigourakou and Efthimios S. Skordas[*]**

*Section of Solid State Physics and Solid Earth Physics Institute, Department of Physics, National and Kapodistrian University of Athens, Panepistimiopolis, Zografos 157 84, Athens, Greece.*

**Abstract**

Various experimental techniques, have revealed that the predominant intrinsic point defects in $BaF_2$ are anion Frenkel defects. Their formation enthalpy and entropy as well as the corresponding parameters for the fluorine vacancy and fluorine interstitial motion have been determined. In addition, low temperature dielectric relaxation measurements in $BaF_2$ doped with uranium leads to the parameters $\tau_0$, E in the Arrhenius relation $\tau=\tau_0 \exp(E/k_B T)$ for the relaxation time $\tau$. For the relaxation peak associated with a single tetravalent uranium, the migration entropy deduced from the pre-exponential factor $\tau_0$, is smaller than the anion Frenkel defect formation entropy by almost two orders of magnitude. We show that, despite their great variation, the defect entropies and enthalpies are interconnected through a model based on anharmonic properties of the bulk material that have been recently studied by employing density-functional theory and density-functional perturbation theory.

---

[*] eskordas@phys.uoa.gr

## I. INTRODUCTION

Despite their simple structure, fluorides exhibit interesting properties and find a wide range of applications. As typical examples, $CaF_2$ and $BaF_2$, are strongly ionic, wide band-gap materials of interest for use in precision vacuum ultraviolet lithography (e.g., see Ref [1] and references therein). A large number of experimental and theoretical studies on their defect properties have been carried out. It has been shown that the dominant point defects are anion Frenkel pairs and ionic transport can be successfully described through migration of anion vacancies and interstitials [2,3].

Because of its simplicity and strongly temperature-dependent effects, the fluorite structure may serve as a model system for anharmonic calculations. A full anharmonic treatment is not available yet, but one approach is the quasiharmonic approximation in which anharmonicity is taken into account through the volume dependence of phonon frequencies. Along these lines, $BaF_2$ was studied in a recent paper [4] as a model fluorite employing the methods of density-functional theory and density-functional perturbation theory. The focus in that work was on volume- and pressure-dependent properties. Elastic constants and their volume and pressure dependence were presented along with the calculation of the anharmonic property thermal expansion. The present paper shows that these anharmonic properties are also closely connected to the thermodynamic parameters for the defect formation and migration processes in $BaF_2$.

In view of the above, the structure of the present paper is as follows: We first summarize, in Section II, the thermodynamic parameters for the anion Frenkel formation and migration deduced from ionic conductivity studies [5-7] in $BaF_2$. Further, we give the defect migration parameters for the dielectric relaxation process that can be extracted from the analysis of the measurements [8] of the real and imaginary part of the dielectric constant of $BaF_2$ doped with uranium. Second, in Section III, we review a model that interconnects the defect formation and migration parameters with bulk properties. This model, in Section IV, is applied to the case of $BaF_2$. A discussion follows in Section V while in Section VI we summarize our conclusions.

## II. EXPERIMENTAL DATA

We start with the defect parameters obtained from ionic conductivity studies. These are given in Table 1 by following the detailed review paper of Bollmann [6]. In particular, the enthalpy, h, for anion Frenkel formation is 1.81eV and the corresponding entropy, s, is 7.85$k_B$ (where $k_B$ is Boltzmann's constant). For fluorine vacancy migration, the corresponding enthalpy is 0.59eV and the entropy is 1.12$k_B$, while for the fluorine interstitial motion the values are: h=0.79eV and s=3.26±049 $k_B$.

We now proceed to the parameters associated with the (re)orientation process of the electric dipoles formed due to the addition of aliovalent impurities. In general, for reasons of charge compensation, in alkali halides for example, for each divalent cation, a cation vacancy is produced. Thus, an electric dipole is formed consisting of a divalent cation and a neighboring cation vacancy [9]. As a second example, in alkaline earth fluorides doped with rare-earth ions, for each trivalent dopant ion a fluorine interstitial is produced. Thus, electric dipoles are formed. The relaxation time τ of these dipoles studied e.g., by the ionic thermocurrents method or by measuring the real ($\varepsilon'$) and imaginary ($\varepsilon''$) parts of the dielectric constant, is given by the Arrhenius relation:

$$\tau = \tau_0 \exp(E/k_B T) \qquad (1)$$

where $E$ stands for the activation energy and $\tau_0$ is a preexponential factor. The relaxation time $\tau$, which is the time needed for an electric dipole to change orientation in space (apart from a numerical factor), is interconnected with the complex dielectric susceptibility $x^*(\omega)$, where $\omega$ denotes the (angular) frequency of the applied electric field. For example, in the well known case of the Debye response, this interconnection reads: $x^*(\omega)/x(0) = (1+i\omega\tau)^{-1}$, where $x(0)$ designates $x(\omega)$ for $\omega \rightarrow 0$. Note that $x^*$ and the complex dielectric permittivity $\varepsilon^* (= \varepsilon' - i\varepsilon'')$ are interconnected through the relation: $\varepsilon^* = \varepsilon_0(1+x^*)$ where $\varepsilon_0$ is the permittivity of the free space.

For a linear $\ln\tau$ versus 1/T plot, the energy $E$ is just the enthalpy h for the corresponding migration (re-orientation) process [10-12]. The pre-exponential factor $\tau_0$ is then interconnected to $s^m$, the corresponding migration entropy, through [12]:

$$s^m = k_B \ln[\tau_0^{-1}/(\lambda \nu)] \qquad (2)$$

where $\lambda$ denotes the number of jump paths accessible to the jumping species with an attempt frequency $\nu$. For example, $\lambda=2$ for "jumps" of the next nearest neighbor interstitial-fluorine ion between equivalent sites.

Montgomery et al. [8] have identified the electrical relaxation peaks of uranium doped BaF$_2$. They concluded that the relaxation having an activation energy of 0.07eV is associated with a single tetravalent uranium ion and the corresponding (pre)exponential factor is $\tau_0$ =3.67×10$^{-12}$s (see Table 1 of Ref. [8]). By inserting this $\tau_0$ value into Eq. (2) and considering that in this case [12] $\lambda$ =4 as well as the approximation [13] $v=v_{T0}$ ($k\rightarrow 0$), where $v_{T0}$ denotes the frequency of the long wavelength transverse optical mode we find s≈0.12k$_B$. Here we used the value $v_{T0}$=0.6×10$^{13}$s$^{-1}$ calculated by Schmalzl [4].

We now plot in Fig. 1 all the entropy values given in Table 1 versus the corresponding experimental enthalpy values for the following four defect processes: anion Frenkel formation, fluorine vacancy migration, fluorine interstitial motion and the relaxation associated with single tetravalent uranium ion. Despite the variation of the relevant enthalpy or entropy values by almost two orders of magnitude, Fig.1 clearly reveals that the entropy increases almost linearly with the enthalpy. It will now be shown that the slope of this plot conforms to an early theormodynamical.

### III. THE MODEL THAT INTERCONNECTS POINT DEFECTS

It has been suggested [14-20], that the defect Gibbs energy g$^i$ is interconnected with the bulk properties of the solid through the relation:

$$g^i = c^i B\Omega \qquad (3)$$

where $c^i$ is a dimensionless constant, B stands for the isothermal bulk modulus, $\Omega$ the mean volume per atom. The superscript *i* refers to the defect process under discussion. This model is usually called cB$\Omega$ model and has been extensively reviewed in Ref. [12]. By differentiating Eq.(3) with respect to temperature *T*, we find the entropy $s^i$ $(=(dg^i/dT)_P)$ and then inserting it into the relation $h^i = g^i + Ts^i$ the corresponding enthalpy $h^i$ is obtained. These two values have a ratio:

$$\frac{s^i}{h^i} = -\frac{\beta B + \frac{dB}{dT}\big|_P}{B - T\beta B - T\frac{dB}{dT}\big|_P} \qquad (4)$$

where $\beta$ stands for the volume thermal expansion coefficient. We emphasize that Eq.(4) states that the ratio $s^i/h^i$ is governed by anharmonic properties of the bulk material, because in an (ideal) harmonic solid we have $\beta = 0$ and $(dB/dT)_p = 0$ [12].

It has been checked that, for different defect processes, e.g. defect formation, defect migration, self-diffusion activation, the cB$\Omega$ model agrees with the experimental data in various categories of solids [14-24] including mixed alkali halides [25-27]. It was also found to satisfactorily describe the defect parameters when applying uniaxial stress variations in a (ionic) crystal [28] (during which electric signals are emitted in a similar fashion to those detected before earthquakes [29-39]).

**IV. APPLICATION OF THE MODEL TO BaF$_2$**

Before proceeding, note that Eq. (4) predicts that the ratio of s/h remains constant for different defect processes in the same host crystal and that is governed by the anharmonic bulk quantities β and dB/dT. Let us now apply Eq. (4) to estimate the ratio s/h in the present case of $BaF_2$. By inserting the experimental values [40]: $β=0.55×10^{-4} K^{-1}$, $B=0.569×10^{11} N/m^2$ and $(d \ln B / dT)_P = -3.9×10^{-4} K^{-1}$, we find: $s/h=3.07×10^{-4} K^{-1}$. This value corresponds to the straight line depicted in Fig.1. Comparing the experimental points with the straight line, we see a satisfactory agreement for the parameters corresponding to the electrical relaxation as well as for those of the fluorine interstitial migration, if the experimental uncertainty for the latter point is also taken into account. Concerning the other two points, i.e., for the fluorine vacancy motion and for the anion Frenkel formation, which seem to deviate markedly from the straight line, their deviation is likely to be smaller for reasons that will be explained in the next Section.

## V. DISCUSSION

In the extrinsic range of the conductivity curve, the slope results in the enthalpy $h^{m,υ}$ for the fluorine vacancy ($υ$) migration ($m$) and the intercept contains the term $ν \exp(s^{m,υ}/k)$ where $ν$ denotes the relevant attempt frequency and $s^{m,υ}$ the entropy for the fluorine vacancy migration. In the so called intrinsic region II, the fluorine vacancies are the more mobile species. Thus, in this region, to a first approximation the slope leads to an activation enthalpy equal to $h^{m,υ} + \frac{1}{2} h^f$, where $h^f$ denotes the anion Frenkel formation ($f$) enthalpy, and the intercept to an activation ($act$) entropy $s^{act}$ equal to $s^{m,υ} + \frac{1}{2} s^f$ where $s^f$ stands for the anion Frenkel formation entropy. (This of course holds to a first approximation, as mentioned, because in practice the conductivity data are computer analyzed using nonlinear least squares techniques to yield a self-consistent best fit of thermodynamic parameters for the defects). Bollmann [6], in order to extract the $s^{m,\upsilon}$ value from the extrinsic range, adopted as usual the approximation $\nu = \nu_D$ where $\nu_D$ is the Debye frequency ($\approx 7 \times 10^{12} s^{-1}$) and then obtained the $s^{m,\upsilon}$ value $s^{m,\upsilon} = 1.12 k_B$. Obviously, if we use instead, the approximation $\nu = \nu_{TO}$, where $\nu_{TO}$ denotes the frequency of the long wavelength transverse optical mode (which seems to be more reasonable for reasons explained in Ref. [12]), which as mentioned above is equal to [4] $\nu_{TO} = 6 \times 10^{12} s^{-1}$, we find $s^{m,\upsilon} = 1.27 k_B$, i.e., a value larger by $0.15 k_B$. This reflects the following: First, the asterisk in Fig. 1, will move closer to the straight line predicted by the $cB\Omega$ model. Second, the $s^f$ value would result smaller by $0.3 k_B$, i.e. one should find the value $s^f = 7.55 k_B$ instead of the value $s^f = 7.85 k_B$ reported by Bollmann [6]. Hence, the triangle in Fig. 1 will also move closer to the straight line obtained from the $cB\Omega$ model. In other words, when one uses in the analysis of the conductivity curve $\ell n(\sigma T)$ vs $1/T$ the approximation $\nu = \nu_D$ (instead of the more reasonable approximation [12] $\nu = \nu_{TO}$, as mentioned), the actual value of $s^{m,\upsilon}$ is underestimated, while $s^f$ is overestimated. This reflects that the scatter of the experimental values around the straight line in Fig. 1 predicted by the $cB\Omega$ model is likely to be smaller. As far as the origin of this linearity is concerned, it solely stems from the extent of the validity of Eq. (3) for different defect processes (e.g., the anion Frenkel formation, and the anion migration). This validity for different defect processes has been discussed in detail in chapter 14 of Ref. [12]. Furthermore, note that as shown in Ref. [14], Eq.(3) which has a thermodynamical origin, is superior to other empirical relations that have been suggested in the past and use, instead of the isothermal bulk modulus B [=$(C_{11}+2C_{12})/3$], a different combination of the elastic constants $C_{ij}$, e.g., the shear modulus $C_{44}$ or $(C_{11}-C_{12})/2$. This has been repeatedly demonstrated in a large variety of solids, e.g., monovacancy formation in fcc- and bcc- metals [16, 18] as well as in rare gas solids [22], formation of Schottky pairs in alkali halides and cation Frenkel pairs in silver halides [10, 15], cation vacancy migration in alkali halides [17], etc.

Finally, we comment on the slope of the line drawn in Fig. 1 predicted by the cBΩ model. Recall that this slope was calculated by using the early elastic and expansivity data reported by Wong and Schuele [40]. We clarify that it was confirmed that this slope is not affected if we use instead, the more recent elastic data by Leger et al. [41] that have been obtained by studying angle-dispersive x-ray and neutron time-of-flight powder diffraction as a function of pressure. In addition, we mention that, since the normal pressure phase of $BaF_2$ (which has the fluorite structure) transforms [41,42] at about 3 GPa to an orthorhombic structure, we extended our investigation to the latter structure and found that the value of the slope changes markedly. Unfortunately, the extent to which this result of the cBΩ model quantitatively agrees with experimental data cannot be checked, because defect entropies and enthalpies have not yet been determined experimentally in the new phase (orthorhombic) of $BaF_2$. Other properties, however, e.g., the luminescence of $Eu^{+2}$-doped fluoride crystals $Ba_xSr_{1-x}F_2$ (x=0, 0.3, 0.5 and 1), have been recently measured [43] and showed that there exists an anomalous pressure dependence

## VI. CONCLUSIONS

The defect entropies and enthalpies for anion Frenkel formation, fluorine vacancy migration and fluorine interstitial motion in $BaF_2$ were investigated. These have been complemented with the relevant parameters extracted from the analysis of the electrical relaxation peak associated with a single tetravalent uranium. Interestingly, the entropies and enthalpies vary by almost two orders of magnitude and reveal a proportionality between the s- and h-values. In addition we find that the ratio of s/h conforms to a theoretical value solely deduced from bulk anharmonic properties, i.e., the thermal volume expansion coefficient and the temperature variation of the elastic constants.

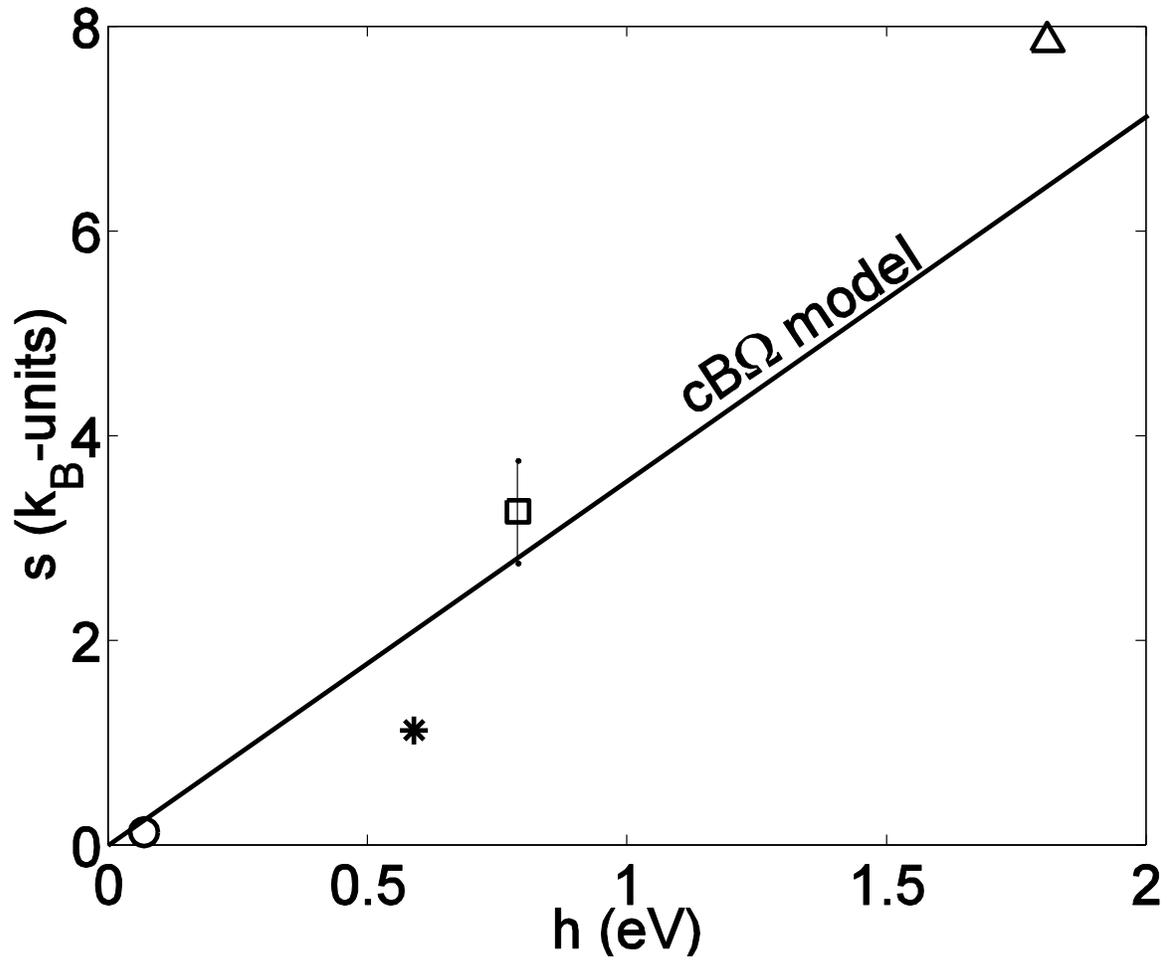

**FIGURE CAPTION**

The defect entropy s versus the defect enthalpy h for four defect processes in $BaF_2$. Triangle: Anion Frenkel formation; Open square: fluorine interstitial migration; asterisk: fluorine vacancy motion; open circle: electrical relaxation associated with a single tetravalent uranium. The straight line is the prediction of the thermodynamic cBΩ model.

TABLE 1. Defect entropies and enthalpies in $BaF_2$

| Process | h (eV) | s ($k_B$-units) |
|---|---|---|
| Fluorine vacancy motion | 0.59 | 1.12 |
| Fluorine interstitial migration | 0.79 | 3.26±0.49 |
| Anion-Frenkel formation | 1.81 | 7.85 |
| Electrical relaxation associated with a single tetravalent uranium | 0.07 | 0.125 |